\documentclass[fleqn,10pt]{wlscirep}
\usepackage{epstopdf}

\title{Observation of turbulence in wave-induced oscillatory flows}

\author[1,2,*,+]{Alberto Alberello}
\author[3,+]{Federico Frascoli}
\author[4,+]{Miguel Onorato}
\author[2,+]{Alessandro Toffoli}
\affil[1]{Centre for Ocean Engineering Science and Technology, Swinburne University of Technology, Hawthorn, VIC 3122, Australia}
\affil[2]{Department of Infrastructure Engineering, The University of Melbourne, Parkville, VIC 3010, Australia}
\affil[3]{Department of Mathematics, Swinburne University of Technology, Hawthorn, VIC 3122, Australia}
\affil[4]{Dipartimento di Fisica, Universit{\`a} di Torino, Via P. Giuria 1, 10125 Torino, Italy}

\affil[*]{alberto.alberello@outlook.com}

\affil[+]{these authors contributed equally to this work}

\keywords{wave motion, oscillatory flow, turbulence, structure functions}

\begin{abstract}

The dynamic and thermal regimes of climate are regulated by an exchange of energy and momentum between the atmosphere and the ocean. The role exerted by surface waves on this interchange is particularly enigmatic. Waves induce turbulence in the upper ocean by breaking and through Langmuir circulations. However, waves can directly inject energy into subsurface layers. This relates to waves not being truly irrotational and therefore the induced orbital motion being turbulent. The existence, extent and properties of this turbulent oscillatory flow still remain uncertain. Here we present measurements of the velocity field of oscillatory flows, which are induced by mechanically generated random wave fields in a large scale experimental facility. Velocities were recorded at a depth sufficiently far from the water surface to rule out effects of wave breaking. We demonstrate that the spectral tail of the velocity field follows a power-law scaling close to $\omega^{-5/3}$. The turbulent behaviour is investigated via rigorous statistical analysis of the structure functions to highlight the emergence of intermittency in oscillatory flows. The results show that wave motion is turbulent and can contribute to ocean mixing. By deepening of the mixed layer, wave induced motion affects cyclogenesis and sediment resuspension.

\end{abstract}
\begin{document}

\flushbottom
\maketitle

\thispagestyle{empty}

\section*{\label{sec:intro}Introduction}

Wind stresses and waves (both surface and internal) generate an irregular and variable motion in the upper ocean layers, commonly referred to as ocean turbulence. This stirs and homogenises water properties (e.g. temperature and salinity) over a depth varying from 25 to 200\,m, which is commonly known as the mixed layer depth. As ocean turbulence regulates heat and momentum exchanges between the atmosphere and ocean, the extent of the mixed layer depth plays a major role in controlling the dynamic and thermal regimes of global climate \cite{csanady2001air,babanin2009wave}. 

In this context, waves play a fascinating role. The wavy surface directly induces an oscillatory motion, with horizontal velocity $u$, throughout a depth comparable to half of the wavelength (under the assumption of infinite water depth). Wave motion is normally described as a potential flow, hence the velocity field is assumed to be irrotational and water particles follow a laminar, almost circular, trajectory. Therefore, waves can inject turbulence into the upper ocean only if they break. As the effect of wave breaking is confined to the most superficial sublayers, within a depth comparable to the wave height ($\approx 10$\,m) \cite{thorpe2007introduction}, the impact on the mixed layer is only marginal. 

In reality, water viscosity is small, but not nil as assumed in potential theory. Accordingly, the velocity field is not truly irrotational \cite{phillips1961note,kinsman1965wind,yefimov1971spectra,yefimov1971wave,teixeira2002onthedistortion}. Provided a wave amplitude based Reynolds number $Re={aU}/{\nu}$\,\cite{babanin2006waveinduced} (with $a$ being the wave amplitude, $U$ the wave orbital velocity and $\nu$ the kinematic viscosity) is large, the wave orbital motion can become unstable and turbulence can, in principle, develop. The peculiarity of a rotational wave motion is that turbulence is injected directly throughout the water column without the need of being advected or diffused. As a result, waves can induce a prompt expansion of an initially shallow mixed layer to depth of the order of $\approx 100$\,m, especially during storms \cite{pleskachevsky2011turbulent,toffoli2012effect}. The recent implementation of parametric models for wave-induced turbulence into global circulation models appears to improve predictions of the mixed layer depth in particular and global climate in general  \cite{babanin2009wave,huang2010wave,huang2011improving,ghantousone2014dimensional}.

Similarly to other natural turbulent flows, Kolmogorov-Obukhov theory \cite{kolmogorov1941dissipation,kolmogorov1941local,obukhov1941distribution} may apply to wave induced turbulence \cite{benilov2012turbulence}. At small scales, for a homogeneous and isotropic flow, away from the boundaries and in the limit of infinite Reynolds numbers, turbulence can be characterised by the scaling properties of the structure functions $S_p$, which are defined as the moments of the distribution of longitudinal velocity increments $\delta u$ between a time interval $\tau$: $S_p(\tau) = \langle|\delta u(\tau)|^p\rangle$. Here $\langle \cdot \rangle$ denotes an ensemble average and $p$ is a positive integer that describes the order of the statistical moment. Note that reference to the time domain is made upon the Taylor's frozen turbulence hypothesis \cite{taylor1922diffusion}. Assuming that turbulence is statistically self-similar, there exists a unique scaling exponent $\zeta_p$ such that $S_p(\tau) \propto \tau^{\zeta_p}$. Following Kolmogorov's four-fifth law \cite{kolmogorov1941dissipation}, $\zeta_p = p / 3$.

There is experimental evidence of deviations from Kolmogorov scaling \cite{frisch1995turbulence,benzi1993extended}. Bursts of activity, associated to the formation of coherent structures, break self-similarity at small $\tau$ leading to anomalous scaling exponents. This phenomenon is known as intermittency.

The determination of the scaling exponents is challenging due to uncertainties related to viscous effects, inhomogeneity, anisotropy, violation of Taylor's hypothesis, and experimental errors \cite{anselmet1984high,frisch1995turbulence}. Whereas low order ($p\leq 4$) structure functions are less prone to uncertainties, errors might become significant at high orders. To evaluate the scaling exponent $\zeta_p$ more accurately, the Extended Self-Similarity (ESS) can be employed \cite{benzi1993extended}. Given the exact solution $\zeta_3=1$ \cite{kolmogorov1941dissipation,kolmogorov1941local}, the other exponents can be derived relatively to $S_3$, i.e. $S_p(\tau)\propto S_3(\tau)^{\zeta_p}$. Validity of the relative scaling (relative to $p=3$) extends beyond the inertial range allowing for a more robust estimation of $\zeta_p$ \cite{benzi1993extended}.

At the second order, the structure function relates to the variance of the spectral density. Therefore, the scaling $S_2 \propto \tau^{2/3}$ implies that the tail of the velocity spectrum decays according to a power law $E(\omega) \propto \omega^{-5/3}$\, \cite{obukhov1941distribution}. In absence of a shear flow, by measuring the wave motion immediately below the trough of non-breaking, mechanically generated, monochromatic waves in a flume, a sporadic behaviour somewhat resembling the $\omega^{-5/3}$ power law in the velocity spectrum has been observed \cite{babanin2009existence}. To a certain extent, results hinted to the existence of wave induced turbulence. This was further substantiated by indirect experimental measurements of wave generated mixing in a thermally stratified fluid \cite{dai2010experiment}. Nevertheless, there is a lack of a rigorous statistical framework in support to these findings, leaving uncertainties on the turbulent properties of wave induce motion \cite{beya2012turbulence}. 

Here we discuss observations of the longitudinal velocity induced by an irregular (random) wave field at an arbitrary depth in absence of wind. Wave are mechanically generated and hence the wave induced motion is the only source of velocity. A large amount of data were collected to build a robust statistical scheme to evaluate the scaling properties of the structure functions and thus verify the presence of turbulence in a wave induced oscillatory flow. In the first Section the laboratory experiments and initial conditions are presented. In the subsequent Section, the spectral tail and the structure functions of the wave induced motion are discussed in details. Final remarks are given in the conclusive Section.

\section*{\label{sec:lab}Laboratory experiment}
\subsection*{\label{sec:lab1}Set-up}

\begin{figure}[htbp]
\centerline{\includegraphics[trim={0 0 0 0},clip,height=0.4\textwidth]{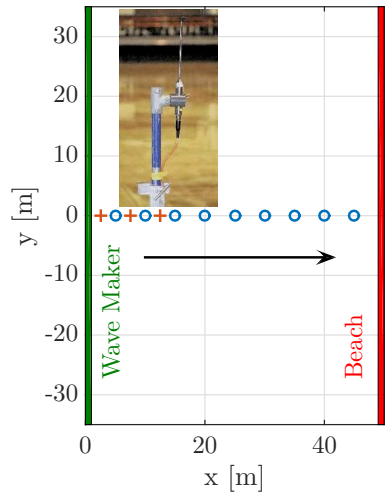}}
		\caption{Schematic of the experimental set-up and example of current-metre in the inset: current-metres (x); and wave probes {o}.}
\label{fig:mtek}
\end{figure}

The experiment consists in monitoring the properties of an oscillatory flow induced by mechanically generated, irregular wave fields. Tests are conducted in the ocean wave basin at MARINTEK (Norway). The facility is $70$\,m wide, $50$\,m long and is equipped with an adjustable bottom \cite{toffoli2011occurrence,toffoli2013experimental}. For the present experiment, a water depth $d= 0.78$\,m is used. A schematic of the experimental set up is shown in Fig.~\ref{fig:mtek}.  

A multi-flap wave-maker is installed on the 70\,m side for the generation of irregular wave field with an arbitrary directional spreading. A second wave-maker is available on the 50\,m side. However, this is not used for the present experiment and hence it simply acts as a vertical wall. Opposite to the wave-maker, sloping beaches are installed to absorb incoming waves. Reflections in amplitude is estimated to be less than 5\% after 30 min of irregular waves of peak period of 1\,s. 
The water surface elevation is monitored at a sampling frequency of 200\,Hz with resistance wave gauges. Probes are deployed along the mean wave direction at progressive distances from the wave-maker. The velocity field is recorded by three electromagnetic current-metres (at 200\,Hz), also aligned with the mean wave directions at 2.5, 7.5 and 12.5\,m from the wave-maker. The current-metres provide the horizontal velocity components along the mean wave direction ($u$) and its orthogonal counterpart ($v$) at a depth of $z=-0.35$\,m.
 
It is important to mention that the instruments are held in position by slender pipes moored at the bottom (see the inset in Fig.~\ref{fig:mtek}). The diameter of the pipes is much smaller than the wavelength. Therefore, turbulent wakes detaching from the structures dissipate rapidly. Their effect on the wave induced motion is thus considered to be negligible \cite{sarkpaya2002experiments}.
 
\subsection*{Wave and velocity field}

Irregular surface waves are generated at the multi-flap wave-maker by imposing an input, directional wave spectrum $E(\omega, \vartheta) = S(\omega) \: D(\vartheta)$ , where $S(\omega)$ is the angular frequency spectrum and $D(\vartheta) $ is the directional spreading function. $S(\omega)$ is modelled with a JONSWAP spectrum \cite{komen1994dynamics} with peak period $T_P=1.64$\,s (corresponding to a wavelength of 3.81\,m), significant wave height $H_S=0.18$\,m and peak enhancement factor $\gamma=6$. This defines a wave steepness $k_P H_S / 2 = 0.16$ (i.e. waves are weakly nonlinear) and a relative water depth is $k_P\:d = 1.29$ (i.e. intermediate water depth). Note that the orbital velocity is recorded at a depth equivalent to $1/10$ of the wavelength and approximately $2$ times the wave height.

A directional function of the form $D(\vartheta) = \text{cos}^N(\vartheta)$, with $N$ being the spreading coefficient, is assigned to model the wave in the directional domain. A narrow directional spreading equivalent to $N=840$ is used. This ensures that all frequency components move along the same direction of propagation, i.e. the wave field is unidirectional. 

The input spectrum is converted into an initial signal by an inverse Fourier transform. In order to have enough samples to produce a statistical analysis, four repetitions are performed by using different sets of random amplitudes and phases. For each test, 20-minute time series are recorded, including the initial ramp-up. At each probe, a total of $3.6\times10^5$ measures of the surface elevation and velocity are gathered.

\begin{figure}[htbp]
\centerline{\includegraphics[height=0.4\textwidth]{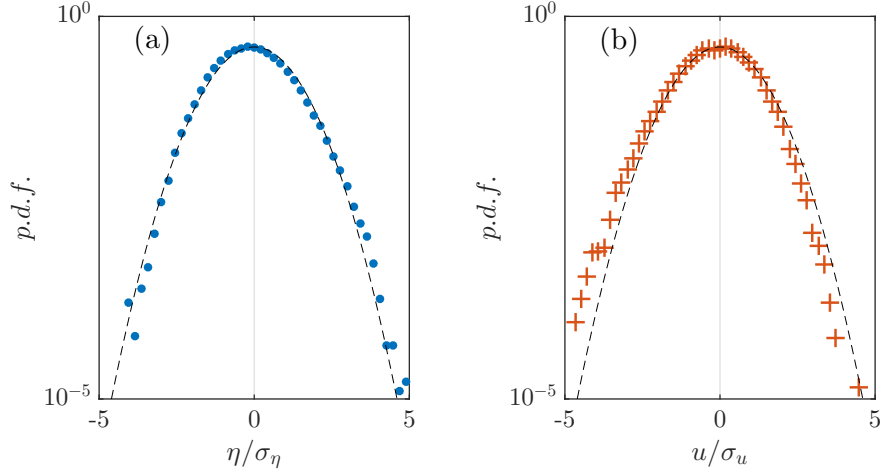}}
		\caption{Probability density function of $(a)$ surface elevation $\eta$, $(b)$ longitudinal orbital velocity $u$. The distributions for a Gaussian random process are also shown as reference.}
\label{fig:pdf}
\end{figure}

To restore the natural variability, amplitudes are randomised using the Rayleigh distribution, while phases are assumed to be uniformly distributed in the interval [0, $2\pi$). Therefore, the initial oscillation is a Gaussian random process. In intermediate water depth, the chosen sea state is weakly nonlinear. The initial Gaussian wave field evolves into a slightly non-Gaussian sea state within the first metres from the wave-maker after which stationary wave conditions are achieved. Under these conditions formation of large waves ($H > 2\:H_S$) does not occur and breaking is not detected at the current metre locations \cite{onorato2009statistical,toffoli2010evolution,toffoli2010maximum}. An example of the probability density function of the surface elevation at about 12.5\,m is presented in Fig.~\ref{fig:pdf}\,a. As a result of wave nonlinearity, there is a concurrent amplification of the orbital velocity. Accordingly, its probability density function also exhibits a clear departure from Gaussian statistics typical of second order nonlinearity (see Fig.~\ref{fig:pdf}\,b). Note that negative velocity dominates the orbital motion at $z=-0.35$\,m\, \cite{alberello2016non}. For further data processing, only the probe at 12.5\,m from the wave-maker is used for the analysis to exclude the transition from Gaussian to nonlinear statistics.

For the dominant wave component of wave period $T_P$ and wave height $H_S$, orbital motion is characterised by a Reynolds number $Re\approx4.2\cdot10^4$ at the water surface. The decay of the velocities in intermediate water depth is slower compared to the exponential decay predicted in deep water conditions \cite{babanin2009existence}, in fact the Reynolds number scales as $[\cosh(k(z+d))/\sinh(kd)]^2$ according to linear wave theory. At $z=-0.35$\,m, where measurements are undertaken, $Re$ falls to about $\approx1.8\cdot10^4$. The Reynolds only describe the mean properties of the flow, in random seas $Re$ can locally increase by four times under the largest waves ($H/H_S>2$). We observe that the Reynolds number used in the experiments is rather low compared to oceanic conditions, we expect that in the ocean the inertial range would be wider.

\section*{\label{sec:results}Results}
\subsection*{\label{sec:rspectra}Velocity spectrum}

\begin{figure}[htbp]
\centerline{\includegraphics[height=0.4\textwidth]{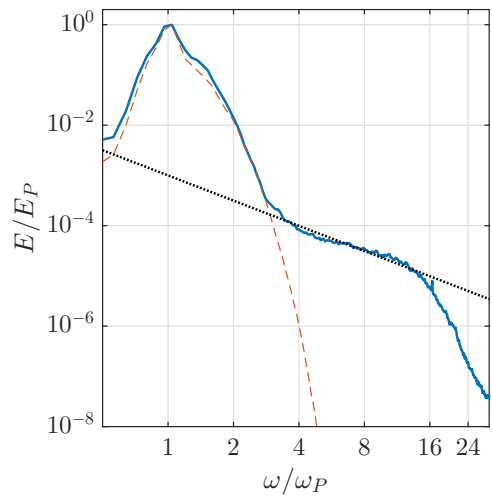}}
		\caption{Averaged dimensionless velocity spectra; the continuous line denotes the spectrum calculated directly from the current-metres, the dashed line denotes the velocity spectrum obtained from surface elevation using Eq.~\ref{eq:eta2u}, the dotted line is the reference slope $-5/3$. Axes are in logarithmic scale.}
\label{fig:sp1}
\end{figure}

The velocity spectrum is calculated from non-overlapping windows of 4096 data points. An ensemble average over all time series and probes is presented in Fig.~\ref{fig:sp1}. A reference spectrum is also shown; this is inferred from the surface elevation by using linear potential wave theory \cite{donelan1992simple}: 
\begin{equation}
    \mathcal{F}\{u_{ref}\}=\mathcal{F}\{\eta\}\cdot\omega\frac{\cosh[(k(z+d)]}{\sinh(kd)},
\label{eq:eta2u}    
\end{equation}
where $u_{ref}$ is the calculated velocity, $g$ the acceleration due to gravity and $\mathcal{F}$ is the Fourier Transform.

The tail of the reference linear spectrum decays rapidly. This is consistent with the fact that high frequency components are in a regime of deep water and dissipate exponentially. Note that the spectral peak is slightly downshifted due to a nonlinear energy transfer from high to low frequency \cite{onorato2009statistical}. The measured spectrum fits the reference one well within the integral range ($\omega < 3\:\omega_P$), i.e. at frequencies associated to the orbital motion. However, an energy cascade occurs at higher frequencies with a consequent departure from the reference spectrum, suggesting the existence of chaotic fluctuations of the dominant orbital motion. There is evidence of an inertial range following the -5/3 power law as predicted by Kolmogorov's theory in agreement with previous experimental observations \cite{babanin2009existence}. The inertial scale, however, remains confined on a very narrow frequency range, spanning between 4 and 15 times $\omega_P$, due to the low Reynolds number \cite{tropea2007springer}. Note that this short inertial scale makes the slope of the spectral tail sensitive to data uncertainty and contamination from the integral and viscous scales \cite{anselmet1984high,frisch1985fully}. Hence corroboration for fully developed turbulence in the oscillatory motion is less firm. For higher frequencies (above 20 times $\omega_P$), the rapid decay of the wave spectrum is evidence of a viscous range, which terminates the energy cascade.

\subsection*{\label{sec:rstru}Structure functions}

\begin{figure}[htbp]
\centerline{\includegraphics[height=0.4\textwidth]{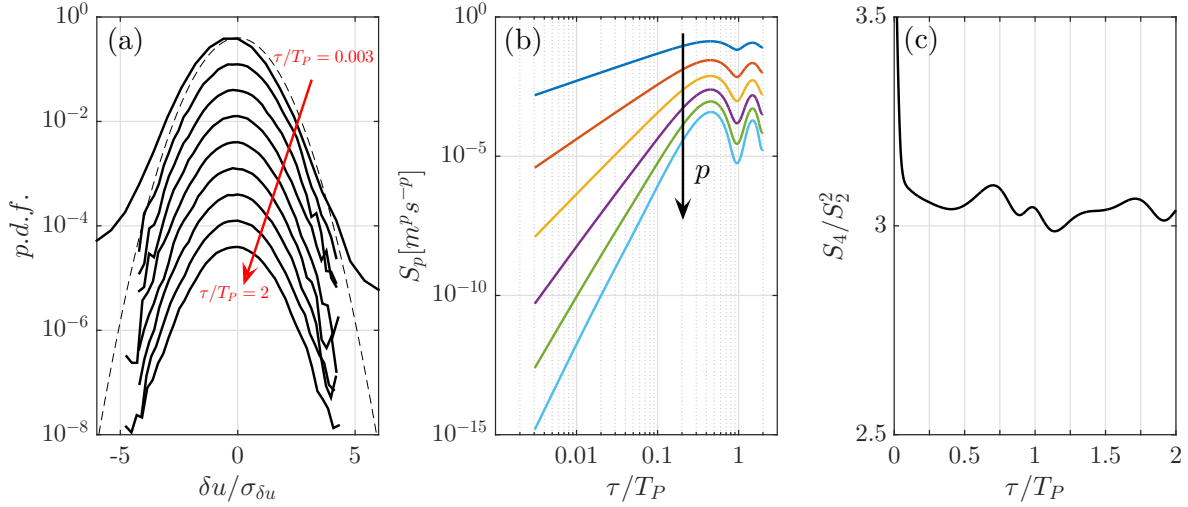}}
		\caption{Statistical properties of the velocity increments. In $(a)$ the probability density function of the velocity increments $\delta u$ for different time separations $\tau$ (increasing in the direction of the arrow) plotted every 1/4$T_P$ (plots are shifted for readability). The distributions for a Gaussian random process is also shown as reference as dashed line. In $(b)$ the structure functions of the first order velocity increments from $p=1$, at the top, to $p=6$, at the bottom. Axes are in logarithmic scale. In $(c)$ the flatness of the velocity increments.}
\label{fig:cef33}
\end{figure}

A more robust analysis of wave induced turbulence is carried out by analysing the structure functions of the velocity increments at small frequency \cite{frisch1995turbulence}. Turbulence is statistically described by the probability density function (p.d.f.) of the velocity increments. Interestingly the p.d.f. modifies its shape for increasing separation distances, see Fig.~\ref{fig:cef33}\,a. At the smallest separation distance, i.e. $\tau$ equal to the sampling frequency the probability density function (p.d.f.) of the velocity increments appears fat-tailed underlying an intermittent behaviour. At higher time separation the p.d.f. tends to Gaussianity after which a periodic widening and opening of the tails can be observed.

Statistical properties of the velocity increments are expressed by the structure functions $S_p(\tau) = \langle|\delta u(\tau)|^p\rangle$. We compute $S_p$ for orders up to $p=12$ and $\tau$ between the sampling frequency and twice the dominant wave period of the oscillatory motion (see Fig.~\ref{fig:cef33}\,b for orders up to $p=6$). The structure functions increase linearly with $\tau$, in a double logarithmic plane, up to $\tau/T_P\approx0.4$. For higher time separations the structure function undergoes oscillations related to the periodicity of the orbital motion. Periodic oscillations of the statistical moments are highlighted by the flatness $S_4/S_2^2$, see Fig.~\ref{fig:cef33}\,c. The flatness, analogously to the kurtosis, highlights deviation from Gaussianity. Flatness is very large for small separations, while it rapidly decreases with increasing $\tau$. For higher $\tau$ the flatness oscillates around 3 which correspond to a Gaussian process. We note that local maxima in the flatness diagram are associated to local minima in the structure function.

Oscillations of the flatness and the structure functions are not detected in classical turbulence experiments; structure functions grow monothonically and flatness tends to Gaussianity for infinite separation times. However, for turbulence associated to a dominant oscillatory flow the maximum separation distance coincides to roughly half of the integral scale, i.e. $T_P/2$. Velocities that are one period apart have high statistical correlation. Subsequent analysis to detect the exponent of the structure functions is hence performed for $\tau/T_P<0.4$ only (i.e. in this range the structure functions are growing and the flatness is decreasing). This range of separation distances is consistent with the analysis of intermittency in weak-wave turbulence\cite{deike2015role}.

\begin{figure}[htbp]
\centerline{\includegraphics[height=0.4\textwidth]{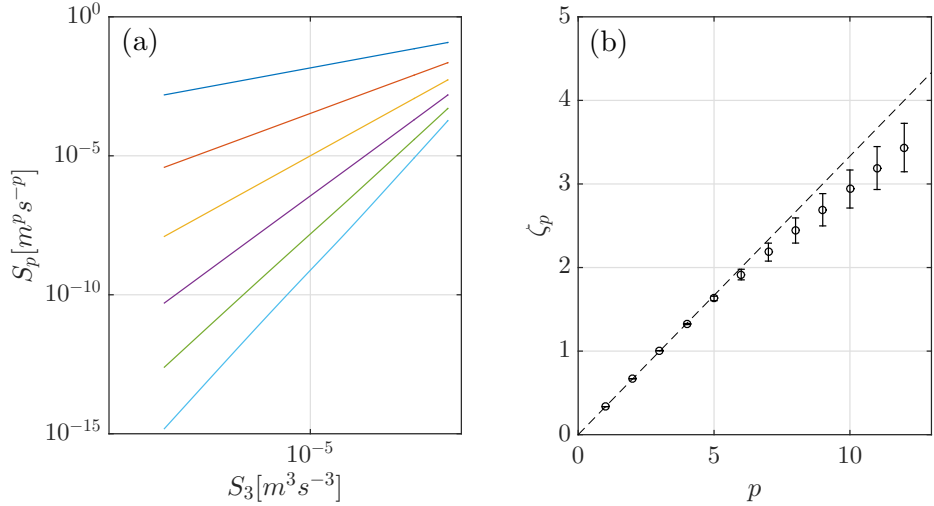}}
		\caption{In $(a)$ structure functions of the first order velocity increments from $p=1$, at the top, to $p=6$, at the bottom plotted against $S_3$. Axes are in logarithmic scale. In $(b)$  relative scaling exponents of the structure function as a function of the order $p$, error-bars show the $\pm \sigma$ confidence interval. The dashed line shows the reference line $\zeta_p=p/3$.}
\label{fig:pf}
\end{figure}

To extract the scaling exponent of the structure functions in a limited inertial range the ESS is applied. Fig.~\ref{fig:pf}\,a highlights how the structure function of order $p^{th}$ linearly relates to the third order structure function. The slope directly provides the relative scaling exponent $\zeta_p$ (we recall that $S_p(\tau)\propto S_3(\tau)^{\zeta_p}$). The $\zeta_p$, i.e. the slopes, are presented in Fig.~\ref{fig:pf}\,b. The value is calculated individually at each probe and then averaged over the number of tests; error bars refer to one time the standard deviation (i.e. 68\% confidence intervals). The exponents $\zeta_p$ fits the Kolmogorov's prediction up to $p=4$, substantiating the turbulent nature of the wave induced velocities (this result is consistent with theoretical findings \cite{benilov2012turbulence}). For $p>4$, experimental data depart from the Kolmogorov's prediction, substantiating that intermittency occurs in wave induced oscillatory flows.

\section*{\label{sec:conc}Conclusions}

Observations of an oscillatory flow forced by mechanically generated, random, unidirectional waves are conducted in a large directional wave basin. A severe storm sea state was applied to obtain large waves and yet minimise the occurrence of wave breaking. Velocity measurements were acquired with electromagnetic current meters at a depth of -0.35\,m, deep enough to exclude contamination due to wave breaking. The current metre was deployed at 12.5\,m from the wave maker. The Reynolds number associated to this sea state is $4.2\cdot10^4$ at the surface and $1.7\cdot10^4$ at the observation depth.

In this condition, the spectral density function of the longitudinal velocity shows evidence of an energy cascade in a narrow inertial range, spanning between $4$ and $15$ times the spectral peak frequency. A power law dependence of the form close to $\omega^{-5/3}$ was detected in agreement with Kolmogorov theory. 

An analysis of the structure functions of the velocity increment was conducted to further verify the robustness of turbulence. The extended self-similarity approach, suitable for low Reynolds number and limited inertial ranges, is adopted to identify the scaling factors. Results show that the structure functions scale according to a power law of the form $\tau^{\:\zeta_p}$, with $\tau$ being the time interval and $\zeta_p$ the exponent of the structure function of order $p$. The exponent is found to fit the Kolmogorov theoretical value of $p/3$ for low statistical moments ($p\leq4$). At higher orders, the exponent departs from the $p/3$ line, corroborating the emergence of intermittency. 

Our experiments suggest that turbulence exists in the ocean further justifying previous working hypothesis. At oceanic scales, i.e. $Re$ one order of magnitude larger, a wider inertial range would develop making the energy cascade even more evident. In limited water depth, due to the slow decay of the velocity with the distance from the surface, wave induced turbulence can extend to the sea bottom affecting sediment resuspension. In deep water conditions, however, the rapid exponential decay of the velocity profile limits the turbulent layer to a depth comparable to half of the wavelength.


\section*{Acknowledgements}

This work was supported by the Swinburne University of Technology Postgraduate Research Award (SUPRA).
Experiments were funded by the E.U. $7^{th}$ Framework Programme through the Integrated Infrastructure Initiative HYDRALAB IV (Contract No. 022441).

\section*{Author contributions statement}

All authors analysed the data and reviewed the manuscript.

\end{document}